\documentclass{aa}
\usepackage{natbib}
\usepackage{txfonts}
\usepackage[dvips]{graphicx}
\usepackage{tabularx}
\usepackage{epsfig}
\usepackage{subfigure}

\bibpunct{(}{)}{;}{a}{}{,}
\usepackage{booktabs,aalongtable}

\begin{document}

\title{A kinematic study of the compact jet in quasar B3 1633+382}

\author{Yi Liu \thanks{e-mail: yliu@shao.ac.cn} \inst{1}, D. R. Jiang
\inst{1}, Zhi-Qiang Shen\inst{1}, M. Karouzos \thanks{Member of the
International Max Planck Research School (IMPRS) for Astronomy and
Astrophysics at the Universities of Bonn and Cologne}\inst{2} }

\institute{Key Laboratory for Research in Galaxies and Cosmology,
Shanghai Astronomical Observatory, Chinese Academy of Sciences, 80
Nandan Road, Shanghai 200030, China \\
\and
Max-Planck-Institut f\"ur
Radioastronomie, Auf dem H\"ugel 69, 53121 Bonn, Germany}

\abstract{

We present a study of the motion of compact jet components in quasar
B3 1633+382. Through analyzing 14 epochs of VLBI observations of
three components (B1, B2, and B3) at 22 GHz, we find two different
possibilities of component classification. Thus two corresponding
kinematical models can be adopted to explain the evolutionary track
of components. One is a linear motion, while another is a helical
model. Future observations are needed to provide new kinematical
constraints for the motion of these components in this source.

 }

\authorrunning{Yi Liu et al.}
\titlerunning{The kinematic study of compact jet in quasar B3 1633+382}

\maketitle

\section{Introduction}
Associated with a redshift of z = 1.814 \citep{Stri1974}, the radio
source B3 1633+382 (4C 38.41) was identified as a quasar
\citep{Paul73} and has been frequently observed at different
wavelengths across the electromagnetic spectrum during the past
decades. A nearly flat spectrum has been derived from 1.5 to 90 GHz
contemporaneous measurements \citep{Land86}. Strong variability at
both long and short radio wavelengths has also been detected
\citep{Span81, Alle92, Seie85, Kuhr81}. This quasar has been
classified as an optically violent variable (OVV)~\citep{Matt93}
because of its strong optical variability. Both infrared and X-ray
fluxes have been obtained with the IRAS and the Einstein Observatory
respectively \citep{Neug86, Ku80}. As one of the most powerful
extragalactic objects detected by the Energetic Gamma Ray Experiment
Telescope (EGRET) on the Compton Gamma Ray Observatory
\citep{Fich94, Thom95, Hart99}, its $\gamma$-ray power is two orders
of magnitude higher than the typical value at any longer wavelength
\citep{Matt93}. This source is also included in the Fermi-LAT
3-month bright AGN list \citep{Abdo09}.

Very Large Array (VLA) observations of B3 1633+382 have shown a
triple structure with an overall size of $12^{''}$ along the
north-south direction. This is a misaligned source, with its VLA jet
at almost 90 degrees position angle to the Very Long Baseline Array
(VLBA) jet \citep{Murp93, Brit07}. A jet in the western direction
with a sharp bend to the south at about 50 mas was detected by
\citep{Pola95} with 1.7 GHz Very Long Baseline Interferometer (VLBI)
observations. Superluminal proper motion $\mu=0.16\pm0.03$ $\rm
mas~yr^{-1}$ (4.4$\pm0.8$~c) was detected on the basis of three
epochs of VLBI observations at 5 GHz \citep{Bart95}. It has also
been observed with the VLBA at dual-frequency \citep{Fey97}. The
VLBI space observatory programme (VSOP) mission, using the HALCA
satellite along with a global array of earth-based telescopes, has
detected a faint core with sub-milliarcsec resolution \citep{List01,
Scot04}. \citealp{Jors001}(2001) presented results of their
multi-frequency monitoring program with the VLBA during six epochs
between 1994 and 1996. A prominent jet at a position angle (P.A.)
$\sim-87^{\circ}$, extending out to 1.5 mas with three components,
has been detected at 22 GHz. The authors found that the proper
motions of components B1 and B3 are 0.20 and 0.14 $\rm mas~yr^{-1}$
(5.5~c and 3.9~c), respectively. Component B2 was shown to be
quasi-stationary at a distance of $\sim0.5$ mas from the core.

In our 22 GHz VLBA observation preformed on 2000 March 1 (see Fig.
1), there are four jet components (in this work, we name them
components B1, B2, and B3 following \citealp{Jors01}) loacted in a
western direction with 0.51, 1.33 and 2.15 mas separation from the
core respectively. Component B4 is the newest one at 0.15 mas from
the core. This result is consistent with another VLBA observation at
43 GHz in 1999 \citep{List00}. However, after combining their proper
motion with \citealp{Jors001}(2001) and ours (see Fig. 2(a)), we
find it hard to distinguish the evolutionary track of the two
internal components owing to the lack of observations from 1997 to
2000. In order to study the kinematic in quasar B3 1633+382, more
observation epochs are needed. Thus we additionally collected and
analyzed seven more VLBA epochs at the same 22 GHz frequency from
the National Radio Astronomy Observatory (NRAO) archive data (see
Table 1). Combined with these seven archive observed data, one
observation of our own, and six other observations obtained from
literature, we have 14 epochs of VLBA observations at 22 GHz for
this source.

We study the kinematics of the compact jet in quasar B3 1633+382.
The paper is organized as follows. In Sect. 2 the data reduction is
described. The motions of components are detailed in Sect. 3,
followed by our conclusions in Sect. 4. We adopt the spectral index
convention $f_{\nu}\propto\nu^{-\alpha}$ and a cosmology with
$H_{0}=70 {~km ~s^ {-1}~Mpc^{-1}}$, $\Omega_{M}=0.3$,
$\Omega_{\Lambda} = 0.7$.

\begin{figure}[h]
\centerline{\epsfxsize=80mm\epsfbox{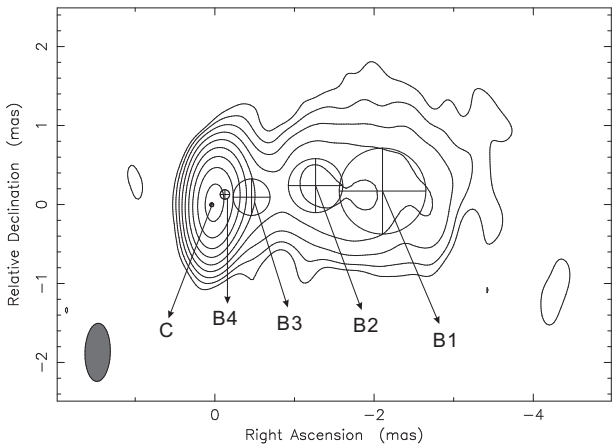}} \caption{Model-fitted
image of the quasar B3 1633+382 at 22 GHz, observed on 2000 March 1.
The restoring beam is 0.737$\times$0.319 mas at P.A. =
-$1.6^{\circ}$. The contour levels are (-1, 1, 2, 4, 8, 16, 32, 64,
128, 256)$~\times$ 2.93 mJy $beam^{-1}$. The peak brightness is
0.992 Jy $beam^{-1}$.}
\end{figure}

\section{Data reduction}\label{vlbiobs}
Our 22 GHz VLBA observation of B3 1633+382 was performed on 2000
March 1. The signals were recorded in four intermediate frequency
(IF) bands for a total bandwidth of 32 MHz with 2 bit sampling. The
recorded data were first correlated at the VLBA correlator in
Socorro (New Mexico) and then calibrated and fringe-fitted using the
NRAO astronomical image processing system (AIPS) software. Initial
amplitude calibration was done using system temperature measurements
and the NRAO-supplied gain curves. The data were corrected for 2 bit
sampling errors and atmospheric opacity. The imaging and
self-calibration were carried out in the DIFMAP package (Shepherd,
Pearson,\citep{Shep94}. Of the seven epochs from the NRAO archived
data, four were observed at 8, 15, 22 and 43 GHz, one at both 22 and
43 GHz, and two at 22 GHz only. All archive data were processed with
the steps described above. However, we encountered a problem with
the correction for atmospheric opacity for the data of 1998 January
2. We used the simultaneously observed 15 GHz data instead to
replace the 22 GHz data in our study, which might introduce a less
than 0.1 mas position shift.

\begin{figure}[h] \centerline{\epsfxsize=120mm\epsfbox{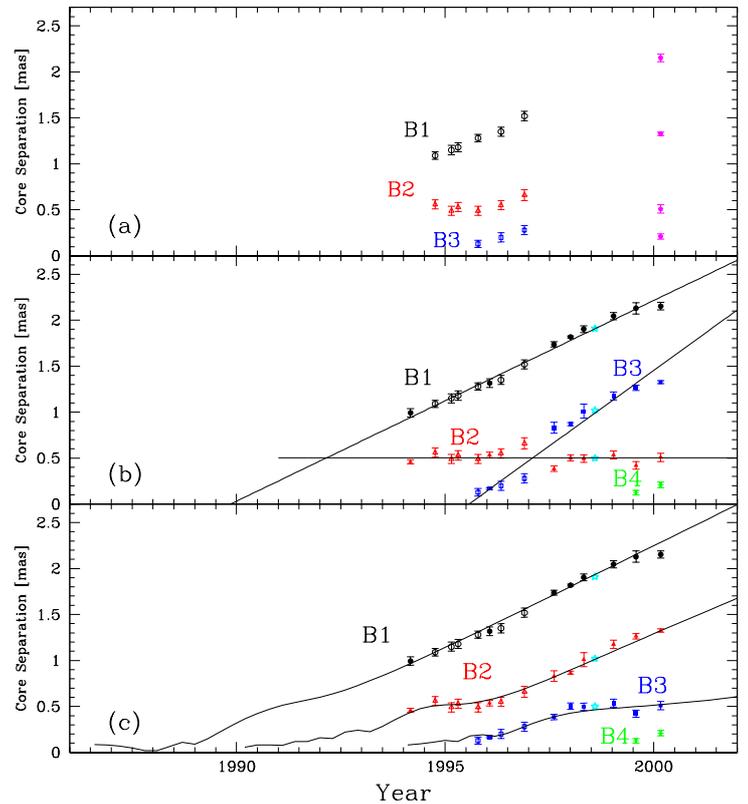}}
\caption{Core separation as a function of time for the individual
jet component. The data combined are from the Jorstad et al. (2001)
(open symbols), archive data, and our observation (solid symbols).
Stars show the components from the VSOP observations by Lister et
al. (2001). The first panel plots the proper motion with Jorstad et
al. (2001) and our own observation (purple pentagon). The second
shows a simple linear proper motion to fit the evolution track of
components of 14 epochs. The last one displays a simple helical fit
to the evolution track of components depending on a different
classification of the components.}
\end{figure}

After obtaining images in DIFMAP, we performed the model fitting to
the self-calibrated visibility data using circular Gaussian
components. For all model fits we assumed the brightest component to
be stationary and used it as a reference point. The other components
were measured relative to it. Uniform weighting of the $u\upsilon$
data was used to ensure the highest resolution. Figure 1 shows the
naturally weighted image of our observation with the model-fitted
circular Gaussian components. During the model-fitting procedure of
components, we found that inner component B3 (see Fig. 1) was
consistently detected in all seven archive data and our own
observation (see Table 1). At the same time, component B3 had about
half as much flux density as core when it was ejected from the core.
Because this was similar to a situation of the innermost component
B4, it was continuously detected in at last two observations.
Component B4 also displays a flux density comparable to that of the
bright core when it is ejected from the core. That means we could
consider those components B3 and B4 as real and rule out the
possibility that they are just noise on the fringes of the strong
core component. Table 1 is a summary of the model-fitting
parameters. We used data from the VSOP observations by
\citealp{List01} for our component identification. A maximum
uncertainty of 15\,\% in the flux density was calculated from the
uncertainties of the amplitude calibration and from the formal
errors of the model fitting. \citealp{Foma89} presented a method to
estimate the position error as $\Delta r=\frac{\sigma\cdot \Theta}{2
S_{\rm P}}$, where $\sigma$ is the residual noise of the image after
the subtraction of the model, $\Theta$ is the full width at half
maximum (FWHM) of the component, and $S_{\rm P}$ is the peak flux
density. However, this formula might underestimate the actual
position error if the peak flux density $\rm S_{P}$ is very high or
the width $\Theta$ of component is small. Therefore we derived our
position error from the formula $\Delta r=\frac{\sigma\cdot
\Theta}{S_{\rm P}}$. Figure 2 presents the core separation (the
radial distance from the core) as a function of time for the
individual model components (open symbols denote the data at 22 GHz
from published literature, solid symbols show our model-fitting
results in Table 1; stars display components from the VSOP survey by
\citealp{List01}, which is used for our component identification),
where circle, triangle, and square symbols denote the B1, B2, and B3
components, respectively.

\section{Kinematic motion of the jet}
After combining all 14 epochs VLBA observations, we found there are
two different possibilities to classify the components. The first
one is related to a simple linear proper motion (see Fig. 2(b)). The
linear proper motion of component B1 is found to be 0.21 $\rm
mas~yr^{-1}$, consistent with \citealp{Jors001} (2001). It is also
roughly consistent with the result of \citealp{Kell004} (2004), who
found a smaller proper motion of 0.15 $\rm mas~yr^{-1}$ (4.1~c) from
a simple linear fit. As for component B3, we found a linear proper
motion of 0.32 $\rm mas~yr^{-1}$, which is about 50\% higher than
that of component B1. Then we found that is always one blob
$\sim0.5$ mas away from the core at different epochs. If those blobs
were responding to the same component, it would mean that component
B2 is quasi-stationary at distances of $\sim0.5$ mas away from the
core, at least during the observation interval. However, we note
that there is an abrupt transition from the year of 1996 September
to 1997 June for both components B2 and B3. Component B2 experiences
a sudden rapid fallback to the core at an apparent velocity of 0.39
$\rm mas~yr^{-1}$ between these six months, while B2 is modeled to
be in a quasi-stationary motion. Meanwhile, component B3 undergoes a
fast acceleration with 0.78 $\rm mas~yr^{-1}$ during the same time
interval. This proper motion is much higher than that fitted from
all observed data points of component B3. However, we note that
because of the rarity of observations, the apparent acceleration
when component B3 passes through or gets ahead of component B2,
could also be a result of the observations' accuracy. More
observations are needed in the future.

Although a linear fit can explain the motion of components, there is
another possibility, depending on the different classification of
the components (see Fig. 2(c)). In this case, component B2 seems to
remain almost stationary at a core separation of $\sim0.5$ mas
during the period 1994-1997. Subsequently component B2 accelerates
to 0.20 $\rm mas~yr^{-1}$ (5.5~c), a velocity comparable to that of
component B1. This kind of accelerated motion has sometime been
interpreted as a change of the angle between the jet direction and
the line of sight \citep{Houg96, Vice96, Homa01, Wang01}. Meanwhile,
component B3 moves with a similar apparent velocity to component B1
before reaching a core separation of $\sim$ 0.5 mas. When component
B3 is $\sim$ 0.5 mas away from the core, its speed is significantly
low. Thus, it is likely that the two inner jet components (B2 and
B3) show a similar proper motion mode; both appear stationary at a
core-separation of $\sim$ 0.5 mas, but when they leave the region of
$\sim$ 0.5 mas  from the core, they both display velocities on the
order of 0.2 $\rm mas~yr^{-1}$ (5.5~c). Then the linear motion is
not enough to explain both the core separation evolution and the
overall path of the components. Therefore we adopt a simple helical
model presented by \citealp{Stef095} (1995) to fit the evolution
track of the components.

In the helical model, the jet can be described by four physical
parameters, the jet's kinematic energy $E_{kin}$, the opening
half-angle $\psi$ of helix (hereafter called opening angle), the
angular momentum $L_{z}$, and the momentum along the jet axis
$P_{z}$. Keeping three of these four physical parameters constant,
four possible cases were studied. We found only one case (where
$E_{kin}$, $L_{z}$, and $\psi$ are conserved) that was physically
plausible for our kinematic study of B3 1633+382. To fully determine
the curved trajectory with the helical model, 10 free parameters are
required (for an illustration see Fig. 1. in \citealp{Stef095}.
These are: three parameters corresponding to the orientation of the
coordinates (the viewing angle between the jet axis and the
observer's line of sight $\theta$; the phase angle of the helix
$\phi_{0}$; the position angle of the jet axis in the sky plane
$\chi$), two parameters describing the initial point of the jet
ejection (the ejecting time $t_{ej}$ and the distance from the jet
axis $r_{0}$), two parameters accounting for the offset of the
observed VLBI-core from the ejection of the helix in right ascension
($\bigtriangleup\alpha$) and declination ($\bigtriangleup\delta$),
and three parameters of conservation (the initial value for the
Lorentz factor $\gamma$ of a component, the angular velocity
$\omega_{0}$, and the opening angle $\psi$). The best parameters
derived from this simple helical fit are listed in Table 2.

There are several constraints derived by using the same methods as
\citealp{Stef095} (1995): the Lorentz factor ($\gamma\gtrsim ~10$),
the opening angle ($\psi\lesssim~5^{\circ}$), the initial radius
($\rm r_{0}\lesssim~0.1~mas$), the time of appearance for component
B3 ($t_{ej}\thickapprox~1995.2$), and the offset ($\rm
\sqrt{\bigtriangleup\alpha^{2}+
\bigtriangleup\delta^{2}}\lesssim~0.15~mas$). We plot the result of
model fits to the observed trajectory for each component in the left
panels of Fig. 3. Points with error-bars are observed data, whereas
the solid lines are from model calculation.

The apparent velocity $\beta_{app}$ is related to the proper motion
$\mu$ by
\begin{equation}
\beta_{app}=\frac{\mu~d_{L}}{c(1+z)}
\end{equation}
\citep{Pear87}, where $d_{L}$ is the luminosity distance, c is the
speed of light, and z is the redshift. Proper motion calculated from
two neighboring data points are assigned to be measured at the
middle of these two observations. With parameters $\gamma$ and
$\theta$  derived from the helical model fitting for each jet
component, we can calculate the  predicted apparent velocity
$\beta_{app}^{'}$ using
\begin{equation}
\beta_{app}^{'}=\frac{\beta\sin\theta}{1-\beta\cos\theta},
\end{equation}
where $\beta=\sqrt{1-\gamma^{-2}}$ is the velocity of the bulk
motion. $\beta_{app}^{'}$ can be compared to the observed
$\beta_{app}$, in order to evaluate the correctness of this fit. In
the right panels of Fig. 3 we show a comparison of the apparent
velocity $\beta_{app}^{'}$, calculated with the helical model, and
the apparent velocity $\beta_{app}$ estimated from observations.
Points with error-bars and solid lines are corresponding to the
observed data and helical model calculation, respectively.

The model parameters of the helical fitting (e.g., $\gamma$,
$\omega$, $\phi$) listed in Table 2 are consistent for the three
components. Moreover, the single dish radio flux monitoring of this
source at 22 GHz, reported by \citep{Tera98,Tera04}, lasting nearly
twenty years from 1981 to 2000, shows three prominent peaks at about
1986.7, 1991.4 and 1995.3 in its light curve \citep{Brit99, Turl99}.
They can be considered as the ejection time $t_{ej}$ for each
corresponding component, assuming that each new component appearance
is accompanied by an outburst in the total intensity. The ejection
times derived by the helical model for these three components are
1986.6, 1990.1, and 1994.1. If component B1 moves along a linear
path, the $t_{ej}$ derived from proper motion extrapolation would be
around 1989.7, three years later than the flaring time 1986.7.
Compared with the extrapolated ejection times from a linear fit, it
seems that the ejection times derived from the helical model are
more consistent.

\begin{figure}[h]
\centerline{\epsfxsize=90mm\epsfbox{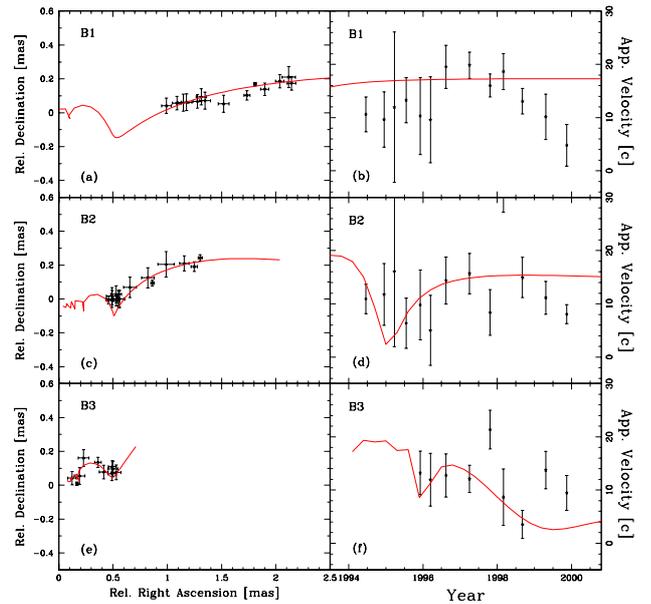}} \caption{Results of
helical model fit to the observed trajectory and the comparisons of
model-calculated apparent velocities $\beta_{app}^{'}$ for each
component (observations shown with black points/black solid line
with points; model shown with red solid line).}
\end{figure}

After combining the 14 epoch observations of this source, we find
the kinematical motions of B3 1633+382 can be reasonably explained
with both a simple linear and helical mode. In the linear
kinematical motion, the component B2 moves at significantly slower
apparent velocity than the other two components in this compact jet.
These low-pattern velocity or even stationary features have
sometimes been explained as standing recollimation shocks in an
initially over-pressurized outflow, which have been reproduced in
numerical simulations of AGN jets (e.g., \citep{Gmez95, Peru07,
List09}). Alternatively, it is possible that the quasi-stationary
motion of the component is due to an extremely small viewing angle
of the specific geometry \citep{Albe00}. We roughly estimate a
Doppler factor of approximately several tens by using both the
average flux density and the average size of the core. Taking the
apparent velocities into account, this implies that the jet is
aligned at $\sim1^{\circ}$ to our line of sight, and the Lorentz
factor corresponding to the pattern velocity is $\rm \Gamma\sim18$.
Although it is an approximate estimation, the direction of the jet
from our line of sight and the Lorentz factor fitted from linear
model are consistent with the parameters from the helical model
calculation. As for the helical fitting, on the other hand, it has
been applied for fitting both the projected trajectory and the
apparent velocity of components in the compact jet of quasar B3
1633+382. The model-calculated projected trajectories also well fit
the significantly low apparent velocity for components B2 and B3
around a core-separation of $\sim$ 0.5 mas. Helical motions combined
with indications of a time-dependent ejection angle can be explained
in the context of a binary core. Binary black hole (BBH) models have
been employed to explain similar component motions in other sources
(e.g., 1803+784; \citep{Brit10, Rola08}, 3C 345; \citep{Loba05}).
The component B3 1633+382 has been shown to exhibit periodicities in
the radio \citep{Fan07} and in the optical \citep{Bozy90}
lightcurves, while also being variable in the $\gamma$-rays
\citep{Fan02}. Quasi-periodic variability across the spectrum is a
further argument in favor of a BBH in the nucleus of this object
(\citealp{Karo10}; e.g., OJ 287; \citealp{Sill88}; \citealp{Leht96},
3C 454.3; \citealp{Qian07}). Helicity and precession of the jet can
be explained in terms of Kelvin-Helmholtz instabilities (e.g.,
\citealp{Came92}, \citealp{Peru06}), precession of the accretion
disk (e.g., \citealp{Capr06}), or magnetic torques \citealp{Lai03}.
Although the stationary proper motion of components in B3 1633+382
can also be explained with both linear and helical model in this
study, the long-term VLBA study has indicated that the standing
components may be in temporarily quiescent states
(\citealp{List09}). Future observations will provide new kinematical
constraints.

\section{Conclusions}
The quasar B3 1633+382 shows two different kinds of proper motion
for its components. We note that it could be explained as two kinds
of kinematical models when we use different classifications for the
components. In the first case, if component B1 and B3 move outwards
with a linear speed, then component B2 stays at a quasi-stationary
state $\sim$ 0.5 mas from the core for the whole observation
interval. Component B1 has a proper motion of 0.21 mas $\rm
yr^{-1}$, which is roughly similar to previous observations. In the
meantime, component B3 also experienes a linear expansion with an
apparent velocity of 0.32 mas $\rm yr^{-1}$. Although this model can
explain the motion of the components, there is another possibility.
In the second case, component B2 was accelerated after remaining
almost stationary at a core-separation of about 0.5 mas for several
years, while component B3 decelerates to a significantly low
velocity at the same core separation of about 0.5 mas. Then we
applied a helical model under the assumption of three conservative
quantities (the jet's kinematic energy, angular momentum, and the
momentum along the jet axis; \citealp{Stef095} 1995) to interpret
the kinematics of the components. The solutions for each component
can be used to explain the projected trajectories and the apparent
velocity with time as well as to demonstrate in particular the low
apparent velocity of the two components B2 and B3. Based on the
predictions from the helical model, the components may have the same
mode of motion. Future observations are needed to provide new
kinematical constraints for the motion of these three components in
this source.

\begin{acknowledgements}
We thank the anonymous referee for insightful comments and
constructive suggestions. We are grateful to Thomas Krichbaum and
Richard Porcas for helpful discussions and proofreading that
improved the presentation of this work. Research Supported by the
CAS/SAFEA International Partnership Program for Creative Research
Teams. This work is supported by NSFC under grants 10803015,
10625314, 10633010, 10821302, 10978009 and supported by the Ministry
of Science and Technology of China (Grant No. 2009CB824900). Yi Liu
also thanks the support from the Knowledge Innovation Program of
Chinese Academy of Sciences. M. Karouzos was supported for this
research through a stipend from the International Max Planck
Research School (IMPRS) for Astronomy and Astrophysics. The VLBA is
operated by the National Radio Astronomy Observatory which is
managed by Associated Universities, Inc., under cooperative
agreement with the National Science Foundation. The National Radio
Astronomy Observatory is a facility of the National Science
Foundation operated under cooperative agreement by Associated
Universities, Inc. This paper has made use of the NASA/IPAC
Extragalactic Database (NED), which is operated by the Jet
Propulsion Laboratory, California Institute of Technology, under
contract with the National Aeronautics and Space Administration.
\end{acknowledgements}

{}

\begin{table*}
 \caption{Gaussian models of VLBI observation.}
\label{table1}
\begin{tabular}{lllllllllll}
\hline\hline

Epoch &   Frequency &Component &Flux & R & $\theta$ & Major \\
       &  (GHz)  &      &(Jy) &(mas)&(deg)&(mas)

\\\hline
  1994.17 & 22.2 & Core & 0.81 $\pm$ 0.12 &  0.0            &  ... &  0.10\\%&  0.977 \\
          &      & B1   & 0.25 $\pm$ 0.04 & 0.99 $\pm$ 0.05 &  -87 &  0.48\\%&   0.013 \\
          &      & B2   & 0.30 $\pm$ 0.05 & 0.46 $\pm$ 0.02 &  -92 &  0.37\\%&   0.026 \\
  1996.07 & 22.2 & Core & 0.74 $\pm$ 0.11 & 0.0             &  ... &  0.03\\%&   9.772 \\
          &      & B1   & 0.28 $\pm$ 0.04 & 1.32 $\pm$ 0.05 &  -87 &  0.78\\%&   0.076 \\
          &      & B2   & 0.08 $\pm$ 0.01 & 0.53 $\pm$ 0.03 &  -98 &  0.30\\%&   0.011 \\
          &      & B3   & 0.46 $\pm$ 0.07 & 0.17 $\pm$ 0.01 &  -86 &  0.21\\%&   0.009 \\
  1997.61 & 22.2 & Core & 1.20 $\pm$ 0.18 &  0.0            &  ... &  0.07\\%&   2.951 \\
          &      & B1   & 0.37 $\pm$ 0.06 & 1.74 $\pm$ 0.03 &  -87 &  0.72\\%&   0.009 \\
          &      & B2   & 0.13 $\pm$ 0.02 & 0.83 $\pm$ 0.06 &  -81 &  0.67\\%&    0.003 \\
          &      & B3   & 0.13 $\pm$ 0.02 & 0.39 $\pm$ 0.03 &  -70 &  0.26\\%&    0.023 \\
  1998.01 & 15.4$^{a}$ & Core & 1.15 $\pm$ 0.17 &  0.0            &  ... &  0.11\\%&    2.344 \\
          &      & B1   & 0.49 $\pm$ 0.07 & 1.82 $\pm$ 0.01 &  -85 &  0.76\\%&    0.021 \\
          &      & B2   & 0.18 $\pm$ 0.03 & 0.87 $\pm$ 0.02 &  -84 &  0.67\\%&    0.010 \\
          &      & B3   & 0.14 $\pm$ 0.02 & 0.50 $\pm$ 0.03 &  -77 &  0.28\\%&    0.045 \\
  1998.32 & 22.2 & Core & 1.38 $\pm$ 0.21 &  0.0            &  ... &  0.07\\%&    3.388 \\
          &      & B1   & 0.35 $\pm$ 0.05 & 1.90 $\pm$ 0.04 &  -86 &  0.77\\%&    0.007 \\
          &      & B2   & 0.15 $\pm$ 0.02 & 1.01 $\pm$ 0.08 &  -78 &  0.74\\%&    0.003 \\
          &      & B3   & 0.20 $\pm$ 0.03 & 0.50 $\pm$ 0.04 &  -82 &  0.33\\%&    0.022 \\
  1999.04 & 22.2 & Core & 1.20 $\pm$ 0.18 &  0.0            &  ... &  0.08\\%&    2.239 \\
          &      & B1   & 0.30 $\pm$ 0.04 & 2.05 $\pm$ 0.04 &  -85 &  0.91\\%&    0.004 \\
          &      & B2   & 0.20 $\pm$ 0.03 & 1.18 $\pm$ 0.04 &  -80 &  0.87\\%&    0.003 \\
          &      & B3   & 0.16 $\pm$ 0.02 & 0.54 $\pm$ 0.04 &  -82 &  0.44\\%&    0.010 \\
  1999.57 & 22.2 & Core & 0.88 $\pm$ 0.13 &  0.0            &  ... &  0.04\\%&    6.607 \\
          &      & B1   & 0.29 $\pm$ 0.04 & 2.13 $\pm$ 0.06 &  -84 &  0.99\\%&    0.004 \\
          &      & B2   & 0.20 $\pm$ 0.03 & 1.26 $\pm$ 0.03 &  -81 &  0.63\\%&    0.006 \\
          &      & B3   & 0.14 $\pm$ 0.02 & 0.42 $\pm$ 0.04 &  -79 &  0.50\\%&    0.007 \\
          &      & B4   & 0.81 $\pm$ 0.12 & 0.13 $\pm$ 0.02 &  -42 &  0.01\\%&    97.724 \\
  2000.16 & 22.2 & Core & 0.82 $\pm$ 0.12 &  0.0            &  ... &  0.05\\%&    3.890 \\
          &      & B1   & 0.25 $\pm$ 0.04 & 2.15 $\pm$ 0.04 &  -85 &  1.09\\%&    0.003 \\
          &      & B2   & 0.15 $\pm$ 0.02 & 1.33 $\pm$ 0.02 &  -79 &  0.68\\%&    0.004 \\
          &      & B3   & 0.07 $\pm$ 0.01 & 0.51 $\pm$ 0.05 &  -79 &  0.46\\%&    0.004 \\
          &      & B4   & 0.33 $\pm$ 0.05 & 0.21 $\pm$ 0.02 &  -52 &  0.12\\%&    0.275 \\
\hline\hline
\end{tabular}
\vskip 0.1 true cm \noindent Note: a denote the correction for
atmospheric opacity for data on 1998 January  encountered\\
a problem. Thus we used the simultaneously observed 15 GHz data to
replace the 22 GHz data.
\end{table*}

\begin{table*}
\caption{Parameters of the helical model.} \label{table2}
\begin{tabular}{lllllllllll}
\hline\hline Component & $t_{ej}$& $\theta$&$\gamma_{0}$&
$\omega_{0}$&$r_{0}$&$\psi$&$\phi_{0}$&$\chi$&$\bigtriangleup\alpha$&$\bigtriangleup\delta$\\
& & $[^{\circ}]$&&$[\frac{^{\circ}}{yr}]$& [ly]
          &$[^{\circ}]$&$[^{\circ}]$&$[^{\circ}]$&[mas]&[mas]\\

\hline
B1& 1986.6$\pm$0.2 & 1.4$\pm$0.2 &  19.8$\pm$1.3 & 15.5$\pm$0.1 &   0.41$\pm$0.7 &  0.21$\pm$0.04 & 21$\pm$6 & 94$\pm$2 &   0.10 & 0.01 \\
B2& 1990.1$\pm$0.2 & 1.0$\pm$0.1 &  19.5$\pm$0.4 & 23.2$\pm$0.1 &   0.30$\pm$0.1 &  0.16$\pm$0.06 &  65$\pm$15& 87.0$\pm$2 &  -0.05 & -0.04 \\
B3& 1994.1$\pm$0.1 & 0.6$\pm$0.1 &  19.5$\pm$0.6 & 20.7$\pm$0.1 &   0.22$\pm$0.1 &  0.12$\pm$0.02 & -79$\pm$13& 73$\pm$5 &  -0.07 & 0.02 \\
\hline\hline
\end{tabular}
\vskip 0.1 true cm \noindent Col. (1): observed components; Col.
(2): time of ejection; Col. (3): angle between the jet axis and the
line of sight; Col. (4): lorentz factor; Col. (5): angular velocity;
Col. (6): initial radius; Col. (7): opening angle; Col. (8): initial
phase angle; Col. (9): projective position angle of jet axis in sky
plane; Col. (10): offset of ejection in right ascension; Col. (11):
offset of ejection in declination.
\end{table*}

\end{document}